\documentclass[12pt,a4paper] {article} 
\usepackage{amsmath,amssymb,epsfig}
\begin{document}

\def \etal  {\hbox{{\it et al.}}}
\def \zphys {{\sl Z. Phys.~}}
\def \NIM   {{\sl Nucl. Instr. Meth.~}}
\def \NP    {Nucl. Phys.~}
\def \PL    {{\sl Phys. Lett.~}}
\def \PR    {{\sl Phys. Rev.~}}
\def \PRL   {{\sl Phys. Rev. Lett.~}}
\def \PRep  {Phys. Reports~}
\def \apgt  {\raisebox{-0.6ex}{$\stackrel{>}{\sim}$}}
\def \aplt  {\raisebox{-0.6ex}{$\stackrel{<}{\sim}$}}
\def \mev   {\,\mathrm MeV}
\def \mevc  {\,\mathrm MeV/c}
\def \mevcc {\,\mathrm MeV/c^2}
\def \gev   {\,\mathrm GeV}
\def \gevc  {\,\mathrm GeV/c}
\def \gevcc {\,\mathrm GeV/c^2}
\def \dr    {$\;^\mid\!\!\!\longrightarrow$}
\def \dz    {\mbox{D$^0$}}
\def \dplus {\mbox{D$^+$}}
\def \ds    {\mbox{D$^{*+}$}}
\def \Ks   {K$^0_{\mathrm S}\,$}
\def \Bbar {$\overline{\mathrm{B}}$}
\begin{titlepage}
\centerline{\large EUROPEAN LABORATORY FOR PARTICLE PHYSICS (CERN)}
\bigskip
\begin{flushright}
December 15, 1998\\
\end{flushright}
\renewcommand{\thefootnote}{\fnsymbol{footnote}}
\begin{minipage}{\linewidth}
\begin{center}
\vspace{20mm}
\begin{LARGE}
{\bf The Signal Estimator Limit Setting Method}\\
\end{LARGE}
\vskip 1.6 cm
        \normalsize {
Shan Jin\footnote
{Department of Physics, University of Wisconsin--Madison,
Madison, WI 53706}$^{,}$\footnotemark[1],
Peter McNamara$^{a,}$\footnotemark[1]}\\
\vspace{0.2cm}
\end{center}
\end{minipage}
\footnotetext{Corresponding address:  CERN/EP Division, 1211
Geneva 23, Switzerland.  Tel: (41 22) 767 7364; fax: (41 22) 782 8370; email: Peter.McNamara@cern.ch, jin@wisconsin.cern.ch.}
\vspace{15mm}
\begin{abstract}
A new method of background subtraction is presented
which uses the concept of a signal estimator to construct a
confidence level which is always conservative and which is
never better than $e^{-s}$.
The new method yields stronger exclusions than the Bayesian method
with a flat prior distribution.
\end{abstract}
\vspace{30mm}
\begin{center}
(Submitted to {\it Nuclear Instruments and Methods A.})
\end{center}
\end{titlepage}

\section{Introduction}

In any search, the presence of standard model background will degrade the sensitivity
of the analysis because it is impossible to unambiguously seperate events
originating from the signal process from the expected background events.
Although it is possible, when setting a limit on a signal hypothesis,
to assume that all observed events come from the signal, a search analyzed
in this way will only be able to exclude signals which are significantly
larger than the background expectation of the analysis.  Background subtraction
is a method of incorporating knowledge of the background expectation into the
interpretation of search results in order to
reduce the impact of Standard Model processes on the sensitivity of
the search.

The end result of an unsuccessful search is an exclusion confidence
for a given signal hypothesis based on the experimental observation.
This confidence level $1-c$ is associated with a signal and background
expectation and an observation, and is required to be conservative.
A conservative confidence level is one in which the False Exclusion
rate, or probability that an experiment with signal will be excluded,
must be less than or equal to $c$, where $c$ is called the confidence
coefficient.

The classical frequentist confidence level is defined such that this
probability is equal to $c$.  In the presence of a sufficiently large
downward fluctuation in the background observation, however, the classical
confidence level can exclude arbitrarily small signals.
Specifically, for sufficiently large background expectations, it is
possible for an observation to exclude the background hypothesis, in which case,
the classical confidence level will also exclude a signal to which the
search is completely insensitive.  In order to prevent this kind of
exclusion, and because there is no ambiguity when zero events are observed,
it is required that all methods must default to a confidence level $1-e^{-s}$
in order to be ``deontologically correct.''  When no events are observed, one should
not perform any background subtraction, and $c$, the probability
of observing zero
signal events should be just $e^{-s}$.  Further, any observation
of one or more candidate events should yield a larger value of $c$.
This correctness requirement
can be easily verified for any method, and any method which is not deontologically
correct should be considered too optimistic.

\section{Bayesian Background Subtraction Method}
A common method of background subtraction\cite{Helene}, based on computing a
Bayesian upper limit on the size of an observed signal given a
flat prior distribution,
calculates the confidence level $1-c$ in terms of the probabilities
that a random repetition of the experiment with the same expectations
would yield a lower number of candidates than the current observation,
which observes $n_{obs}$.  This method computes the background
subtracted confidence to be
\begin{equation}
CL = 1 - c = 1 - \frac{\mathcal{P}(\mathnormal{n_{s+b} \leq n_{obs}})}
                 {\mathcal{P}(\mathnormal{n_{b}   \leq n_{obs}})}
\end{equation}
where $\mathcal{P}(\mathnormal{n_{s+b} \leq n_{obs}})$ is the probability
that an experiment with signal
expectation $s$ and background expectation $b$ yields an equal or
lower number of candidates than the current observation, and
${\mathcal{P}(\mathnormal{n_{b} \leq n_{obs}})}$ is the probability that an
experiment with background
expectation $b$ yields an equal or lower number of candidates than the
current observation.

When $\mathnormal{n_{obs}}$ is zero, this method reduces to
$e^{-s}$, demonstrating that it is deontologically correct.
Further, the probability
of observing $n_{obs}$ events or fewer is equal to
$\mathcal{P}(\mathnormal{n_{s+b} \leq n_{obs}})$,
and the confidence coefficient for that observation is strictly larger than the
probability of observing the result, so this method is conservative.

The method can be extended\cite{LEPHiggs}
to incorporate discriminating variables such as
the reconstructed mass or neural network output values by constructing
a test-statistic $\epsilon$ for the experiment
which is some function of those discriminating variables,
and constructing the confidence level as the ratio of probabilities
\begin{equation}
CL = 1 - c =  1 - \frac{\mathcal{P}(\mathnormal{\epsilon_{s+b} \leq \epsilon_{obs}})}
                       {\mathcal{P}(\mathnormal{\epsilon_{b}   \leq \epsilon_{obs}})}.
\end{equation}
where  $\mathcal{P}(\mathnormal{\epsilon_{s+b} \leq \epsilon_{obs}})$
is the probability that an independent experiment with signal expectation
$s$, background expectation $b$, and some given distributions of discriminating
variables yields a value of $\epsilon$ less than or equal to $\epsilon_{obs}$
seen in the current experiment, and
$\mathcal{P}(\mathnormal{\epsilon_{b} \leq \epsilon_{obs}})$ is
the probability that an independent experiment with background
expectation $b$ and some given distributions of discriminating variables
yields a value of $\epsilon$ less than $\epsilon_{obs}$ seen in the
current experiment.
If the test-statistic is the number of observed events, this method reduces
to the method described above,
though the test-statistic can be constructed as a likelihood ratio
or in some other appropriate way such that larger values of $\epsilon$ are
more consistent with the observation of a signal than lower values.

For an observation of zero events the probabilities
$\mathcal{P}(\mathnormal{\epsilon_{s+b} \leq \epsilon_{obs}})$ and
$\mathcal{P}(\mathnormal{\epsilon_{b} \leq \epsilon_{obs}})$
are simply the Poisson probabilities of observing zero
events in the two cases.  Because a correctly defined test-statistic
has its smallest value when and only when there are no events
observed, the confidence level for the generalized version of this
method then reduces to the same value as the number counting method
when there are no events observed,
and it is deontologically correct.  Similarly, the probability of observing
a more signal-like test-statistic value is equal to
$\mathcal{P}(\mathnormal{\epsilon_{s+b} \leq \epsilon_{obs}})$,
and as $\mathcal{P}(\mathnormal{\epsilon_{b} \leq \epsilon_{obs}}) \leq 1$,
$c$ is always greater than or equal to this value, so the method is
conservative.

\section{Signal Estimator Method}
Though the Bayesian method described in Section~2 satisfies the
criteria set out in Section~1, it is not the only background subtraction
method which is both conservative and deontologically correct.  The
Signal Estimator method satisfies both of these criteria using
$\mathcal{P}(\mathnormal{\epsilon_{s+b} \leq \epsilon_{obs}})$
and a boundary condition to calculate the confidence level.  The boundary
condition imposes the correctness requirement on the confidence level, while
also making the result conservative.

We wish to determine if a given signal hypothesis
$s$ is excluded.  If we could know the observed
test-statistic based on events truly from signal only,
which we refer to as the signal estimator
$(\epsilon_{s})_{obs}$,
the confidence level would be rigorously defined as
\begin{equation}
CL \equiv 1 - c \equiv 1 - \mathcal{P}(\epsilon_{s} \leq (\epsilon_{s})_{obs})
\end{equation}
where $\mathcal{P}(\epsilon_{s}\leq(\epsilon_{s})_{obs})$ is the probability
that an experiment with signal expectation $s$
yields a value of the signal estimator less than or equal to $(\epsilon_{s})_{obs}$.

Unfortunately, we cannot directly know $(\epsilon_{s})_{obs}$ from
an experiment as it is not possible to unambiguously determine if an
event comes from signal or background.  We can only directly know a
test-statistic value based on the total observation
\begin{equation}
\epsilon_{obs} = (\epsilon_{s+b})_{obs}.
\end{equation}
Although it is not possible to know $(\epsilon_{s})_{obs}$ directly,
it is still possible to produce an estimate of it, with which we can
calculate Eq.~3.  This is most straightforward for test-statistics
of the form
\begin{equation}
\epsilon_{s+b} = \epsilon_{s} \oplus \epsilon_{b}
\end{equation}
where `$\oplus$' represents a sum or product.  For example, in simple
event counting,
\begin{align}
\epsilon       & = n \\
n_{s+b}        & = n_{s} + n_{b}.
\end{align}
In this case, we can use a Monte Carlo simulation of the background
expectation to remove the background contribution in
the observed test-statistic $(\epsilon_{s+b})_{obs}$, i.e., to estimate
$(\epsilon_{s})_{obs}$, and to calculate Eq.~3.
In each Monte Carlo experiment, the estimate of $(\epsilon_{s})_{obs}$
is defined as
\begin{equation}
(\epsilon_{s})_{obs} = 
   \begin{cases}
      \epsilon_{obs} \ominus \epsilon_{b} & {\mathrm if} \epsilon_{obs} \ominus \epsilon_{b} \geq (\epsilon_{s})_{min} \\
      (\epsilon_{s})_{min}         & {\mathrm if} \epsilon_{obs} \ominus \epsilon_{b} \le  (\epsilon_{s})_{min}
   \end{cases}
\end{equation}
where `$\ominus$' represents difference or division, and
$(\epsilon_{s})_{min}$ is the minimum possible value of the signal estimator,
which corresponds to the physical boundary (zero signal events).

The confidence level can be computed with Monte Carlo methods in
the following way for an observed test-statistic $\epsilon_{obs}$.
First, generate a set a Monte Carlo experiments with test-statistic
values distributed as for experiments with the expected background
but no signal to determine a distribution of possible signal estimator
values for the observation according to Eq.~8.
Next, using a sample of Monte Carlo with test-statistics distributed
as for experiments with
signal only, and for each possible signal estimator value, calculate
\begin{equation}
c(\epsilon_{obs},\epsilon_{b}) = \mathcal{P}(\epsilon_{s} \leq max[\epsilon_{obs} \ominus \epsilon_{b},
(\epsilon_{s})_{min}]).
\end{equation}
The value of $c(\epsilon_{obs},\epsilon_{b})$ averaged over all of the
signal estimator values determined with
background Monte Carlo forms an estimate of
$\mathcal{P}(\epsilon_{s} \leq (\epsilon_{s})_{obs})$, or
\begin{equation}
c \equiv \mathcal{P}(\epsilon_{s} \leq (\epsilon_{s})_{obs}) \approx
\overline{c(\epsilon_{obs},\epsilon_{b})}.
\end{equation}

The Monte Carlo procedure described above is very slow, and without
generalization, it can only be used for the class of test-statistics
which satisfy Eq.~5.
The method can be generalized into a much simpler mathematical format
which can be used for any kind of test-statistic.  The generalization can
best be illustrated with an example.  In the case of simple
event counting, the boundary condition for the signal estimator
can be understood intuitively.  For
an observation of $\mathnormal{n_{obs}}$ events, the confidence level
is computed by allowing
the background to vary freely, and according to Eq.~8, the signal
estimator will be

\begin{equation}
(n_{s})_{obs} = 
  \begin{cases}
    n_{obs}-n_{b}& \text{if $n_{obs}-n_{b} \geq 0$}\\
    0&             \text{if $n_{obs}-n_{b} \le  0$.}
  \end{cases}
\end{equation}

Using Eq.~10, one can easily compute the confidence coefficient to be

\begin{equation}
\begin{split}
c        &= [ \mathcal{P}(n_{b}=0)       \times \mathcal{P}(n_{s} \leq n_{obs})   \\
         &   +   \mathcal{P}(n_{b}=1)    \times \mathcal{P}(n_{s} \leq n_{obs}-1) +\ldots \\
         &   +   \mathcal{P}(n_{b}=m)    \times \mathcal{P}(n_{s} \leq n_{obs}-m) +\ldots \\
         &   +   \mathcal{P}(n_{b}=n_{obs}) \times \mathcal{P}(n_{s} \leq 0)] \\
         &+   \mathcal{P}(n_{b}\ge n_{obs}) \times \mathcal{P}(n_{s} \leq 0)  \\
         &=   \mathcal{P}(n_{s+b} \leq n_{obs}) + 
           [1-\mathcal{P}(n_{b}\leq n_{obs})] \times e^{-s}.
\end{split}
\end{equation}
This probability reduces to $e^{-(s+b)}+(1-e^{-b})e^{-s} = e^{-s}$
when one observes no candidates, so
it is deontologically correct, and because the confidence level is always
strictly greater than
$\mathcal{P}(\mathnormal{n_{s+b} \leq n_{obs}})$,
it is conservative.

In order to compare the performances of this method with the Bayesian
method, the confidence levels for a simple experiment
are analyzed in Fig.~1.  For this example, the analysis
is assumed to expect three events from a possible signal, and three
events from Standard Model background processes.  For both
methods, when zero events are observed, the confidence level reduces
to $e^{-s}$ while for observations of more events, the signal
estimator method yields a lower confidence coefficient, and thus
a better exclusion confidence level.  For large numbers
of events, $\mathcal{P}(\mathnormal{n_{b} \leq n_{obs}})$ approaches
one, meaning that both methods approach the classical confidence
level and give very similar results.

This method can then be generalized, as the method described in Section~2
was generalized, to include discriminating variables.  The natural
generalization takes the form

\begin{equation}
c = \mathcal{P}(\epsilon_{s+b} \leq \epsilon_{obs}) +
    [1-\mathcal{P}(\epsilon_{b} \leq \epsilon_{obs})]\times e^{-s}.
\end{equation}

For an observation of zero events, the generalized method continues
to give a confidence
level $e^{-s}$, and the confidence level computed with this method is
always conservative, with $c$ strictly greater than
$\mathcal{P}(\mathnormal{\epsilon_{s+b} \leq \epsilon_{obs}})$.

Generating Monte Carlo experiments based on a simplified
Higgs analysis, one can compare the performances of the generalized
Bayesian method described in Section~2 and the Signal Estimator method.
For the comparison it is
assumed that there are three events expected from background
processes,
with mass distributed uniformly between 70 and 90 GeV/$c^2$, and that
the signal process would yield three events, with mass
distributed according to a single Gaussian whose width is 2.5 GeV/$c^2$
centered at 80 GeV/$c^2$.  Using the test-statistic described in
ref.~\cite{Janot}, Fig.~\ref{comp} shows the relative
improvement in confidence level for this experiment.  The Signal
Estimator method is seen to never a worse confidence level
than the generalized Bayesian method.  For an observation of zero candidates,
and for very signal-like observations (as
$\mathcal{P}(\mathnormal{\epsilon_{b} \leq \epsilon_{obs}})$
approaches one) the methods converge.  In the region in between these
extremes, the Signal Estimator method gives confidence levels
up to 20\% better than the generalized Bayesian method while remaining
conservative.

\section*{Conclusion}

More than one method of calculating background subtraction
confidence levels which is
conservative and deontologically correct exist.  The Signal
Estimator method proposed here yields less conservative limits than the
Bayesian method, which should result in an increase in search senstitivity,
giving better limits in unsuccessful searches.

\newpage

\begin{figure}[htbp]
\hspace{0.0\textwidth} 
\mbox{\epsfig{file=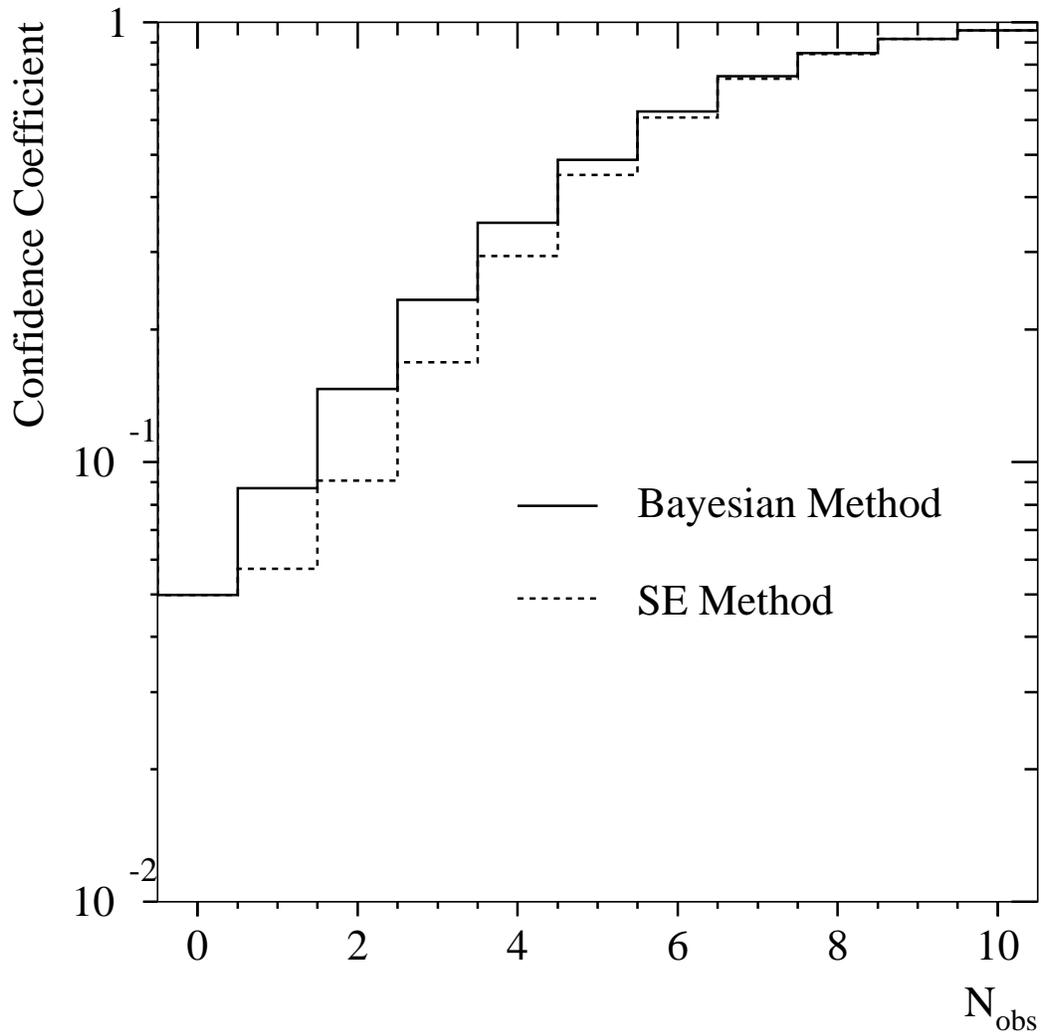,width=0.95\textwidth}}
\caption{A comparison of Signal Estimator method performance to
the Bayesian method performance.  For an experiment with three signal
and three background events expected, the confidence levels are
shown for different numbers of observed events.  The Signal
Estimator method gives either an equal or better confidence level
for all possible observations.}
\label{count}
\end{figure}

\begin{figure}[htbp]
\hspace{0.0\textwidth} 
\mbox{\epsfig{file=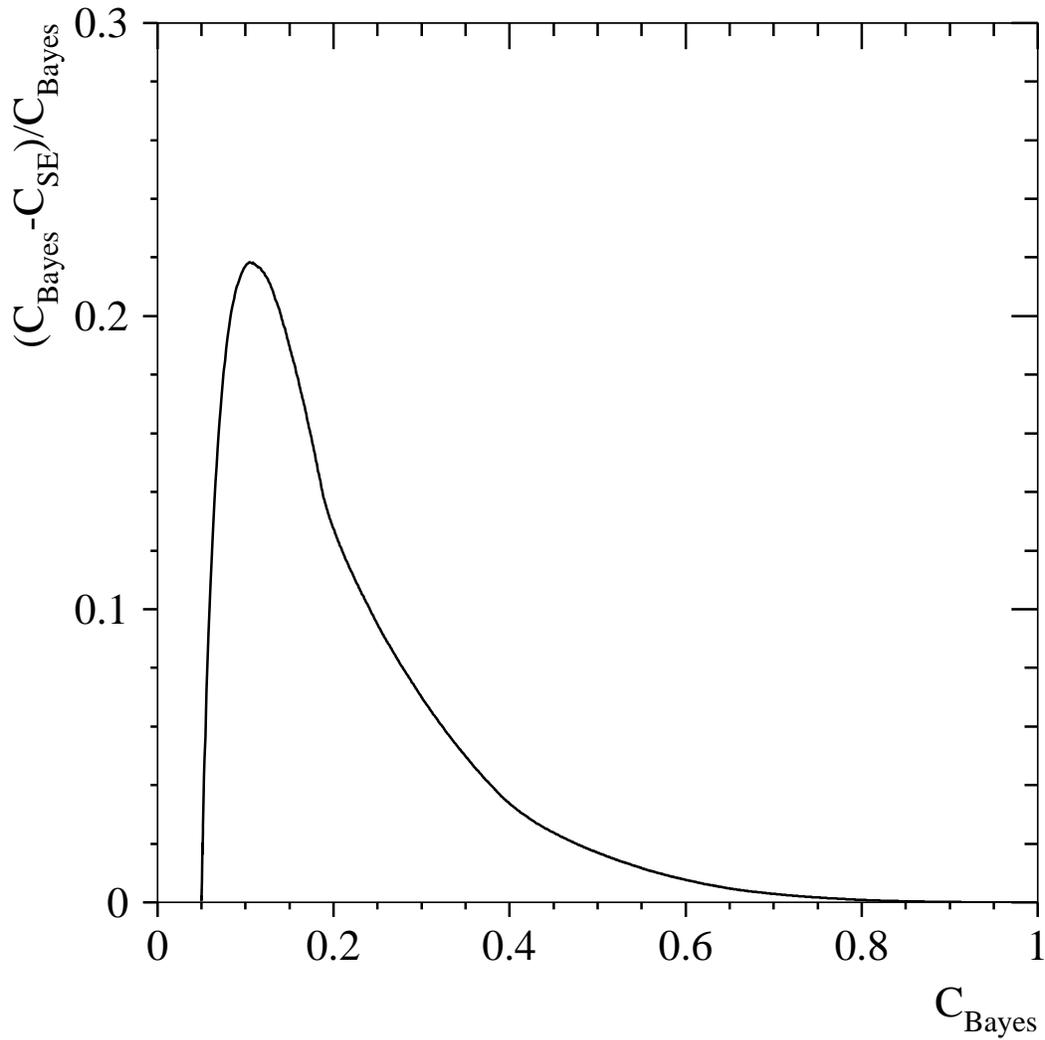,width=0.95\textwidth}}
\caption{A comparison of Signal Estimator method performance to
the Bayesian method performance when discriminating variables are used.
The Monte Carlo experiments assume three signal and three background
events are expected, and the
single discriminating variable has a Gaussian distribution
with width 2.5 GeV/$c^{2}$ for signal, flat for background over a range
of 20 GeV/$c^{2}$.
The relative improvement
in confidence level using the Signal Estimator method is shown
for different confidence level values.}
\label{comp}
\end{figure}

\end{document}